\documentstyle[12pt,epsfig,wrapfig]{article}
\begin{document}
%%%%%%%%%%%%%%%%%%%%%%%%%%% packages... %%%%%%%%%%%%%%%%%%%%%%%%%%%%%%
\def\DAF{DA$\Phi$NE}   
\makeatletter
\def\maketitle{\par
 \begingroup
   \def\thefootnote{\fnsymbol{footnote}}
   \def\@makefnmark{\hbox   
       to 0pt{$^{\@thefnmark}$\hss}}   
   \if@twocolumn               
     \twocolumn[\@maketitle]   
     \else \newpage
     \global\@topnum\z@        % Prevents figures from going at top of page.
     \@maketitle \fi\thispagestyle{empty}\@thanks 
 \endgroup
 \setcounter{footnote}{0}
 \let\maketitle\relax
 \let\@maketitle\relax
 \gdef\@thanks{}\gdef\@author{}\gdef\@title{}\let\thanks\relax}

\def\@maketitle{\newpage
 \null
 \vbox to 10cm{                 %space limit for title/author
 \vskip 1em                    % Vertical space above title.
 \begin{center}
  {\normalsize\bf \@title \par}     % Title set in \normalsize. 
  \vskip 0.5em                % Vertical space after title.
  {\normalsize                % each author set in \normalsize,\bf in a
   \lineskip .5em             % tabular environment
   \begin{tabular}[t]{c}\@author 
   \end{tabular}\par}                   
  \vskip 1em              % Vertical space after author.
\end{center}
 \par
 \vskip 0.5em}                % Vertical space after title/author lines
 }                      % end of vbox to 10cm
\def\abstract{\if@twocolumn
\section*{Abstract}
\else \normalsize
\begin{center}
{ \vspace{-1.0em} {\bf Abstract} \vspace{-.5em}\vspace{0pt}} 
\end{center}
%\quotation 
\fi}
\def\endabstract{\if@twocolumn\else \par\noindent \fi}

% Collapse citation numbers to ranges.  
\newcount\@tempcntc
\def\@citex[#1]#2{\if@filesw\immediate\write\@auxout{\string\citation{#2}}\fi
  \@tempcnta\z@\@tempcntb\m@ne\def\@citea{}\@cite{\@for\@citeb:=#2\do
    {\@ifundefined
       {b@\@citeb}{\@citeo\@tempcntb\m@ne\@citea
        \def\@citea{,\penalty\@m\ }{\bf ?}\@warning
       {Citation `\@citeb' on page \thepage \space undefined}}%
    {\setbox\z@\hbox{\global\@tempcntc0\csname b@\@citeb\endcsname\relax}%
     \ifnum\@tempcntc=\z@ \@citeo\@tempcntb\m@ne
       \@citea\def\@citea{,\penalty\@m}
       \hbox{\csname b@\@citeb\endcsname}%
     \else
      \advance\@tempcntb\@ne
      \ifnum\@tempcntb=\@tempcntc
      \else\advance\@tempcntb\m@ne\@citeo
      \@tempcnta\@tempcntc\@tempcntb\@tempcntc\fi\fi}}\@citeo}{#1}}
\def\@citeo{\ifnum\@tempcnta>\@tempcntb\else\@citea
  \def\@citea{,\penalty\@m}%
  \ifnum\@tempcnta=\@tempcntb\the\@tempcnta\else
   {\advance\@tempcnta\@ne\ifnum\@tempcnta=\@tempcntb \else
\def\@citea{--}\fi
    \advance\@tempcnta\m@ne\the\@tempcnta\@citea\the\@tempcntb}\fi\fi}

%*** below here, standard LaTeX art12.sty with modifications indicated

% art12.sty 23 Sep 85
\lineskip 1pt \normallineskip 1pt
\def\baselinestretch{1}

\def\@normalsize{\@setsize\normalsize{15pt}\xiipt\@xiipt
\abovedisplayskip 12pt plus3pt minus7pt\belowdisplayskip \abovedisplayskip
\abovedisplayshortskip \z@ plus3pt\belowdisplayshortskip 6.5pt plus3.5pt
minus3pt}
\def\small{\@setsize\small{13.6pt}\xipt\@xipt
\abovedisplayskip 11pt plus3pt minus6pt\belowdisplayskip \abovedisplayskip
\abovedisplayshortskip \z@ plus3pt\belowdisplayshortskip 6.5pt plus3.5pt
minus3pt
\def\@listi{\parsep 4.5pt plus 2pt minus 1pt
 \itemsep \parsep
 \topsep 9pt plus 3pt minus 5pt}}
\def\footnotesize{\@setsize\footnotesize{12pt}\xpt\@xpt
\abovedisplayskip 10pt plus2pt minus5pt\belowdisplayskip \abovedisplayskip
\abovedisplayshortskip \z@ plus3pt\belowdisplayshortskip 6pt plus3pt minus3pt
\def\@listi{\topsep 6pt plus 2pt minus 2pt\parsep 3pt plus 2pt minus 1pt
\itemsep \parsep}}
\def\tiny{\@setsize\tiny{7pt}\vipt\@vipt}
\def\scriptsize{\@setsize\scriptsize{9.5pt}\viiipt\@viiipt}
\def\large{\@setsize\large{18pt}\xivpt\@xivpt}
\def\Large{\@setsize\Large{22pt}\xviipt\@xviipt}
\def\LARGE{\@setsize\LARGE{25pt}\xxpt\@xxpt}
\def\huge{\@setsize\huge{30pt}\xxvpt\@xxvpt}
\let\Huge=\huge
\normalsize 
\if@twoside \oddsidemargin 21pt \evensidemargin 59pt \marginparwidth 85pt
\else \oddsidemargin 0.46cm \evensidemargin 0.46cm
 \marginparwidth 68pt 
\fi
\marginparsep 10pt 
\topmargin 0.81cm \headheight 0pt \headsep 0pt \footheight 0pt 
\footskip 35pt 

%*** changed text sizes
\textheight 22cm \textwidth 15cm \columnsep 10pt \columnseprule 0pt 

\footnotesep 8.4pt 
\skip\footins 10.8pt plus 4pt minus 2pt 
\floatsep 14pt plus 2pt minus 4pt \textfloatsep 20pt plus 2pt minus 4pt
\intextsep 14pt plus 4pt minus 4pt \@maxsep 20pt \dblfloatsep 14pt plus 2pt
minus 4pt \dbltextfloatsep 20pt plus 2pt minus 4pt \@dblmaxsep 20pt 
\@fptop 0pt plus 1fil \@fpsep 10pt plus 2fil \@fpbot 0pt plus 1fil 
\@dblfptop 0pt plus 1fil \@dblfpsep 10pt plus 2fil \@dblfpbot 0pt plus 1fil
\marginparpush 7pt 

%***change indents to 1.00cm

\parskip 0pt plus 1pt \parindent 1.00cm  \topsep 10pt plus 4pt minus 6pt
\partopsep 3pt plus 2pt minus 2pt \itemsep 5pt plus 2.5pt minus 1pt 
\@lowpenalty 51 \@medpenalty 151 \@highpenalty 301 
\@beginparpenalty -\@lowpenalty \@endparpenalty -\@lowpenalty \@itempenalty
-\@lowpenalty 
\def\part{\par \addvspace{4ex} \@afterindentfalse \secdef\@part\@spart} 
\def\@part[#1]#2{\ifnum \c@secnumdepth >\m@ne \refstepcounter{part}
\addcontentsline{toc}{part}{\thepart \hspace{1em}#1}\else
\addcontentsline{toc}{part}{#1}\fi { \parindent 0pt \raggedright 
 \ifnum \c@secnumdepth >\m@ne \normalsize\bf Part \thepart \par\nobreak \fi 
\normalsize\bf #2\markboth{}{}\par } \nobreak \vskip 3ex \@afterheading } 
\def\@spart#1{{\parindent 0pt \raggedright 
 \normalsize \bf 
 #1\par} \nobreak \vskip 3ex \@afterheading } 
\def\section{\@startsection {section}{1}{\z@}{-3.5ex plus -1ex minus 
 -.2ex}{2.3ex plus .2ex}{\normalsize\bf}} 
\def\subsection{\@startsection{subsection}{2}{\z@}{-3.25ex plus -1ex minus 
 -.2ex}{1.5ex plus .2ex}{\normalsize\bf}}
\def\subsubsection{\@startsection{subsubsection}{3}{\z@}{-3.25ex plus
 -1ex minus -.2ex}{1.5ex plus .2ex}{\normalsize\sl}}

\setcounter{secnumdepth}{3}

\def\appendix{\par
 \setcounter{section}{0}
 \setcounter{subsection}{0}
 \def\thesection{\Alph{section}}}

\leftmargini 2.5em
\leftmarginii 2.2em \leftmarginiii 1.87em \leftmarginiv 1.7em \leftmarginv 1em
\leftmarginvi 1em
\leftmargin\leftmargini
\labelwidth\leftmargini\advance\labelwidth-\labelsep
\labelsep .5em
\parsep 5pt plus 2.5pt minus 1pt
\def\@listi{\leftmargin\leftmargini}
\def\@listii{\leftmargin\leftmarginii
 \labelwidth\leftmarginii\advance\labelwidth-\labelsep
 \topsep 5pt plus 2.5pt minus 1pt
 \parsep 2.5pt plus 1pt minus 1pt
 \itemsep \parsep}
\def\@listiii{\leftmargin\leftmarginiii
 \labelwidth\leftmarginiii\advance\labelwidth-\labelsep
 \topsep 2.5pt plus 1pt minus 1pt 
 \parsep \z@ \partopsep 1pt plus 0pt minus 1pt
 \itemsep \topsep}
\def\@listiv{\leftmargin\leftmarginiv
 \labelwidth\leftmarginiv\advance\labelwidth-\labelsep}
\def\@listv{\leftmargin\leftmarginv
 \labelwidth\leftmarginv\advance\labelwidth-\labelsep}
\def\@listvi{\leftmargin\leftmarginvi
 \labelwidth\leftmarginvi\advance\labelwidth-\labelsep}
\makeatother
%%%%%%%%%%%%%%%%%%%%%%%%%%% begin... %%%%%%%%%%%%%%%%%%%%%%%%%%%%%%
\pagestyle{plain}
\renewcommand{\rmdefault}{ptm}
\newcommand{\Header}{
  \begin{tabular}{rl}
  \hspace{-.4cm}\includegraphics{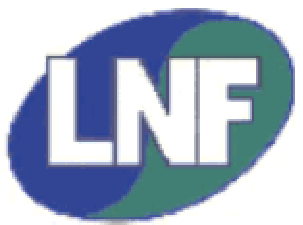} &
    \renewcommand{\arraystretch}{0.5}
    \begin{tabular}{r}
      {\hspace{1cm}~\LARGE\sffamily LABORATORI~ NAZIONALI~ DI~ FRASCATI}\\
      \\
      {\Large\sffamily SIS-Pubblicazioni}\\
    \end{tabular}
    \renewcommand{\arraystretch}{1}
  \end{tabular}
  \vskip 1cm
  \begin{flushright}
  \renewcommand{\arraystretch}{0.5}
    \begin{tabular}{r}
      {\underline{LNF-04/24(P)}}\\    % insert here the preprint number
      {\small 15 November 2004} \\     % insert here the preprint Date
      \\
    \end{tabular}
  \end{flushright}
  \renewcommand{\arraystretch}{1}
  \vskip 1 cm
  }
%%%%%

\setlength{\parindent}{0pt}
\begin{titlepage}
\title{
  \Header
  {\large\bf A scintillating-fiber beam profile monitor for the DAFNE BTF}
}

\vglue -4mm
\author{M. Anelli, B. Buonomo, G. Mazzitelli and P. Valente\\
\it Laboratori Nazionali di Frascati dell'INFN, Frascati, Italy.}
%%%%%%%%%%%%%%%%%%%%%%%%%%%%%%%%%%%%%%%%%%%%%%%%%%%
\maketitle

\begin{abstract}
A scintillating-fiber beam profile detector has been designed, built
and tested, for the monitoring of the position and size of the electron
beam of the \DAF\ BTF, the recently commissioned electron beam-test
facility at the Frascati LNF.
A description of the detector construction and assembly, together with 
the results achieved during the 2003-2004 runs, are here reported. 
\end{abstract}
\vspace*{\stretch{2}}
\begin{flushleft}
% insert here the PACS number 
\end{flushleft}
\begin{center}
%Submitted to Elsevier for pubblication on Nucl. Instrum. Meth. A.
\end{center}

\end{titlepage}

\pagestyle{plain}
\setcounter{page}2
\baselineskip=17pt

%\renewcommand\baselinestretch{1.5}\normalsize

%%%%%%%%%%%%%%%%%%%%%%%%%%%%%%%%%%%%%%%%%%%%%%%%%%%%%%%%%%%%%%%%%%%%%%%%%%%%%%%%%%%%%%%%%%%%%%
%%%%%%%%%%%%%%%%%%%%%%%%%%%%%%%%%%%%%%%%%%%%%%%%%%%%%%%%%%%%%%%%%%%%%%%%%%%%%%%%%%%%%%%%%%%%%%

\section{Introduction}

The \DAF\ Beam-test facility (BTF) provides electron/positron beams with a  
well-defined number of particles in a wide range of multiplicity
and energy, mainly for detector calibration purposes. It was
commissioned in 2002 and it has delivered beam to user experiments
during all years 2003-2004~\cite{bib:btfnote,bib:btfnim}.

A number of detectors have been used for the beam diagnostics 
during the commissioning phase and during the users running periods;
these detectors were mainly intended for the measurement of the number of
particles in the beam, in the full operational range of energy, between 
25 and 750 MeV, and multiplicity, from single particle to $10^{10}$ particles/bunch.

However, a very important point for the operation of the facility is the
measurement of the beam spot position and size, in all the different 
multiplicity and energy configurations and with non-destructive detectors.
High-fluorescence metallic flags are not sensitive at very low beam intensities 
down to the single electron mode, so that position sensitive particle detectors should be used.

For this purpose, a scintillating-fiber detector has been designed, built and tested
during the years 2003-2004 BTF running period. The dector is described in Sec.~\ref{sec:design}
together with some details on the construction and on the readout and
acquisition system, while some experimental results with
the BTF electron/positron beam are reported in Sec.~\ref{sec:results}.

\section{Detector design and construction}
\label{sec:design}

\subsection{Design considerations}
Taking into account cladded scintillating fibers, such as
Pol.Hi.Tech type 0046, a light yield of 3-4 photoelectrons/mm (pe)
at 0.5 m photocathode distance has been measured in the past~\cite{bib:nimprotokloe}. 
The light yield depends on the quantum efficiency of the photomultiplier
and on the quality of the optical coupling with the fibers. In order to have a few
pe per incident particle even without optimal coupling to the photocathode 
(e.g. without employing optical grease), a few layers of fibers have to be stacked one
over the other. This should give a sufficient light yield already with only 
one electron crossing the detector. 
Moreover, by properly staggering the layers, a multi-layer detector allows
to minimize dead spaces between the fibers.

Considering a 4 layers detector of 1 mm fibers, a light yield of 1-2 pe/particle
at the photocathode can be conservatively estimated.
For a typical photomultiplier gain in the range $10^6$-$10^7$ a charge signal of 
0.1-1 pC/particle can then be estimated; this should allow both to be sensitive 
to single electrons, and to reach the 100-1000 particles/bunch range without 
saturation, for a typical 12-bit, 0.25 pC/count, charge integrating ADC.

For a given size of the detector, the number and the area of photocathodes determines the  
the number of readout channels and the number of fibers to be bundled together.
Since the typical BTF beam spot has a Gaussian distribution with $\sigma_x$
$\simeq$ $\sigma_y$ $\simeq$ 5 mm, a two-view detector should have a size of 
at least $5\times5$ cm$^2$, and a millimetric spatial resolution.

A detector consisting of two planes of four layers of 48 fibers of 1 mm diameter
with a readout pitch of 3 mm, will need only 32 channels, and should still be
capable of a millimetric spatial resolution. A total depth of 8 mm of scintillating
fibers corresponds to $\approx 0.02 X_0$, thus giving an almost negligible effect
on the beam energy and spot.

\subsection{Scintillating fibers}
\begin{figure}[tbhp]
  \centerline{\epsfig{file=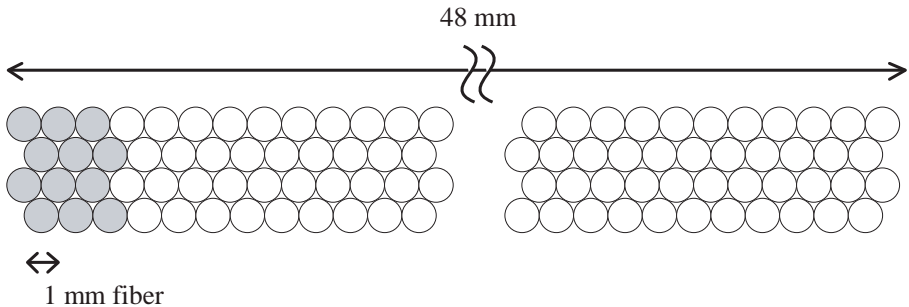,width=0.8\textwidth} }
        \caption{Scheme of the scintillating fibers detector layout: four layers
          of 48 fibers are stacked together (with a 0.5 mm stagger).}
        \label{fig:fibers}
\end{figure}

A two view detector has been constructed, consisting of two identical modules
to be mounted with the fibers running at 90$^\circ$. A single module consists 
of four layers of 1 mm diameter cladded scintillating fibers,
Pol.Hi.Tech type 0046.
The 1 m long fibers have been cut in two $\sim50$ cm pieces, then four
stacked layers of 48 fibers glued side-by-side have been assembled, staggered by
0.5 mm, as shown in the scheme of Fig.~\ref{fig:fibers}.

\begin{figure}[tbhp]
  \centerline{\epsfig{file=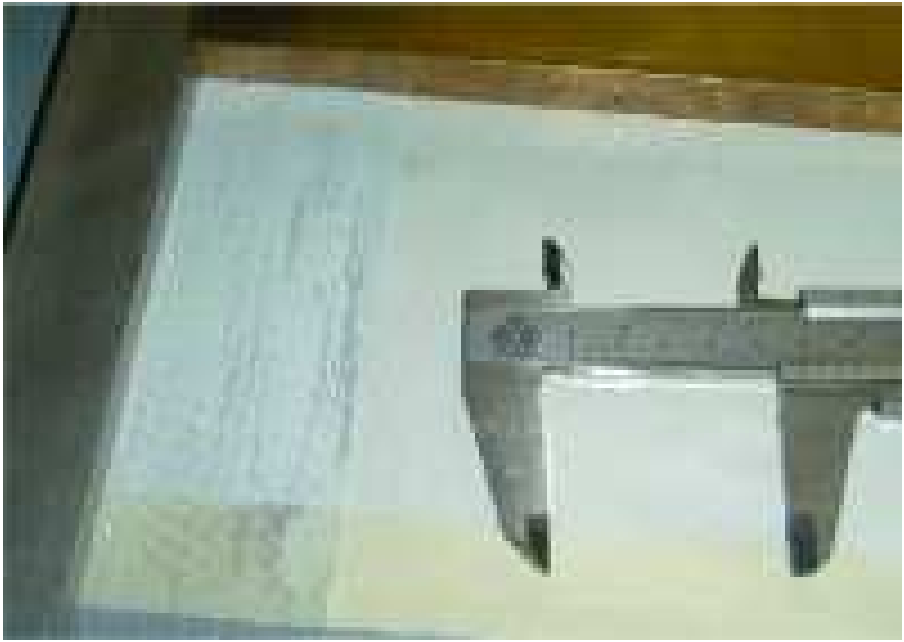,width=0.8\textwidth} }
        \caption{Gluing one layer of 48 scintillating fibers.}
        \label{fig:valentom1}
\end{figure}

The fibers have been glued together by means of optical glue, the BICRON
BC600 Optical Cement, a two component\footnote{We have mixed 100 parts by weight of resin 
and 28 parts by weight of hardener.} clear epoxy resin specifically developed 
for optical joints with plastic scintillators.
A photograph of the first layer of 48 scintillating fibers glued together is
shown in Fig.~\ref{fig:valentom1}; the typical hardening time for each glued layer
was 48 hours.

\begin{figure}[tbhp]
  \centerline{\epsfig{file=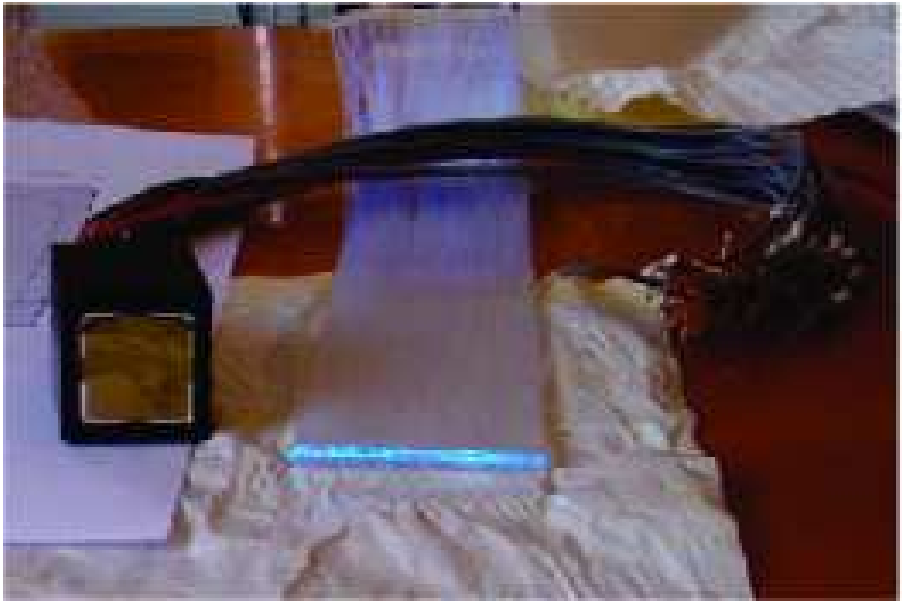,width=0.8\textwidth} }
        \caption{The four layers of scintillating fibers after the gluing (on the right), 
          showed together with the multi-anode PMT (on the left).}
        \label{fig:valentom2}
\end{figure}

In Fig.~\ref{fig:valentom2} a photograph of the four layers of 
scintillating fibers
are shown after the gluing, and before wrapping them in a thin foil
of aluminum. In the same photograph the photomultiplier used for the
readout is also shown (see in the following).

\subsection{Multi-anode photomultiplier}
\label{sec:mapmt}
We have chosen the Hamamatsu H6568 multi-anode photomultiplier (MAPMT) metal package,
based on the R5900-00-M16 tube with a photocathode segmented in 16 pixel of 
$4.2\times4.2$ mm$^2$ each; the package also includes the voltage divider 
circuit. This MAPMT has a good gain uniformity, low 
crosstalk with neightboring channels (below $1\%$) and good timing
performances (1 ns rise time, 0.3 ns FWHM transit time spread).

We usually set the high voltage to -750 V, corresponding to a gain
of $\approx 2\times10^6$.

A group of three fibers for each of the four layers have been bundled together
to cover the area of a MAPMT pixel (such a group of 12 fibers is represented by 
the shaded circles in the schematic view of Fig.~\ref{fig:fibers}).

\begin{figure}[tbhp]
  \centerline{\epsfig{file=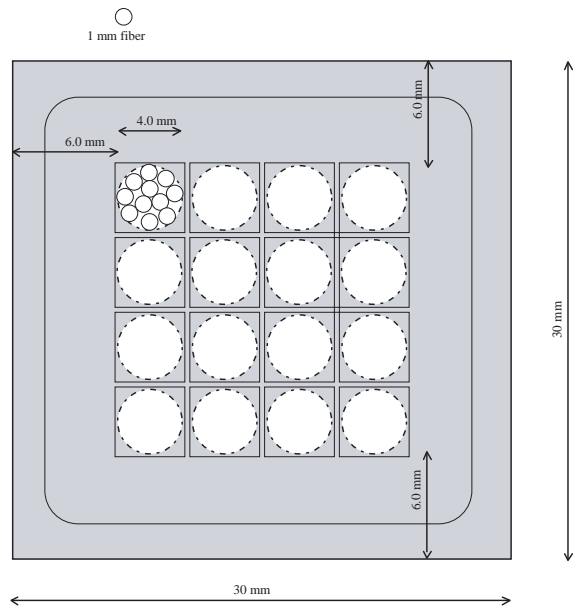,width=0.5\textwidth} }
        \caption{Drawing of the mask for the scintillating fibers 
          bundles, for the coupling to the photocathode surface. 
          For this $4\times48$ scintillating fibers detector the 
          16 bundles consist of 12 fibers. The fibers fit a $\approx$
          4 mm side square, while only 11 fit a 4 diameter circle.
          For sake of construction semplicity, a PVC mask with
          circular holes has been realized, so that only 11/12 fibers
          per bundle are coupled to each MAPMT pixel.}
        \label{fig:valentomf}
\end{figure}

\begin{figure}[tbhp]
  \centerline{\epsfig{file=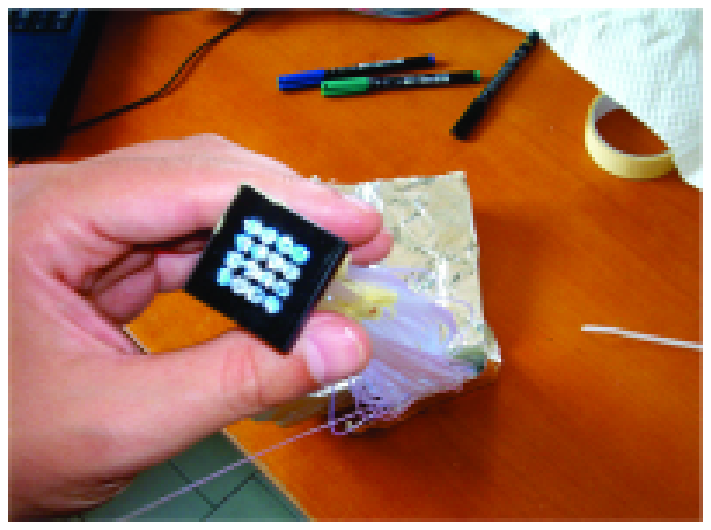,width=0.8\textwidth} }
        \caption{The PVC mask with the 16 scintillating fibers bundles inserted and ready to 
          be coupled to the Hamamatsu H6568 MAPMT.}
        \label{fig:valentom3}
\end{figure}

\subsection{Detector assembly}
For each of the two detector planes, the 16 bundles, constituted by 12 
scintillating fibers, have been inserted in a grooved PVC mask, fitting
the dimensions of the photomultiplier package ($\sim$30$\times$30
mm$^2$), in order to couple each bundle to a pixel on the MAPMT surface.
Since the area of a single pixel is of 4.2$\times$4.2 mm$^2$,
the PVC mask should have been grooved with $\approx$4 mm side square holes,
in order to fit all the fibers in a bundle. However, for sake of construction
semplicity and mechanical robustness, the mask has been grooved with 4 mm diameter
\textit{circular} holes, so that only 11 out of 12 fibers per bundle are actually
coupled to each MAPMT pixel. This is not a problem, both from 
the point of view of the uniformity and of the total yield of light.

A drawing of the mask is shown in Fig.~\ref{fig:valentomf}, while in 
Fig.~\ref{fig:valentom3} a photograph of the PVC mask with the 16 scintillating fibers bundles, 
inserted and ready to be coupled to the MAPMT, is shown.

\begin{figure}[tbhp]
  \centerline{\epsfig{file=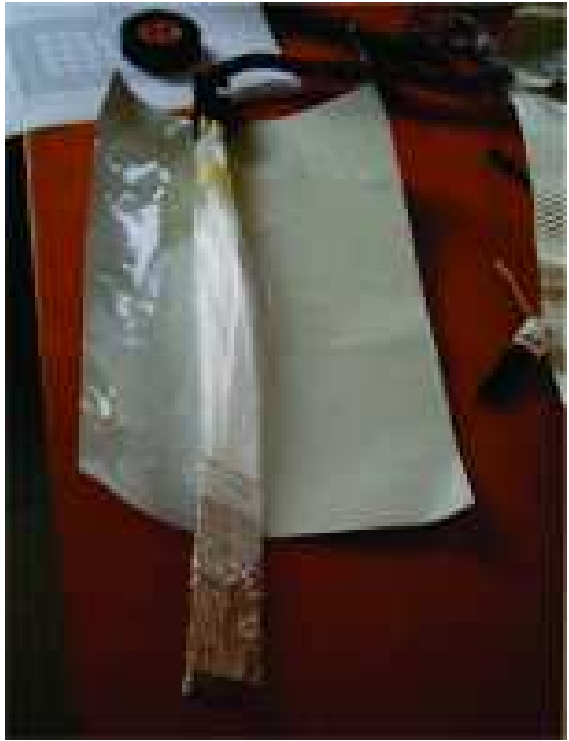,width=0.5\textwidth} }
        \caption{The fiber detector being wrapped with an aluminum foil.}
        \label{fig:valentom4}
\end{figure}

The two planes of the detector have been then wrapped in aluminum foils,
as shown in Fig.~\ref{fig:valentom4}, and finally they have been mounted 
at 90$^\circ$, by means of a mechanical support
as shown in Fig.~\ref{fig:valentom2d}.

\begin{figure}[tbhp]
  \centerline{\epsfig{file=sostegno.ps,width=0.8\textwidth} }
        \caption{The two views of the fiber detector assembled together
          with the fibers running horizontally ($y$ view) and vertically ($x$ view)
          in the plane perpendicular to the beam direction.}
        \label{fig:valentom2d}
\end{figure}

\subsection{Layout on the beam}
During the first test runs, only one plane of the detector was installed,
at the BTF beam exit (at $45^\circ$) usually dedicated to the user setups, 
just before the AIRFLY fluorescence chamber,
with the fibers running in the vertical direction, for measuring the $x$ profile.

\begin{figure}[hbtp]
\centerline{\epsfig{file=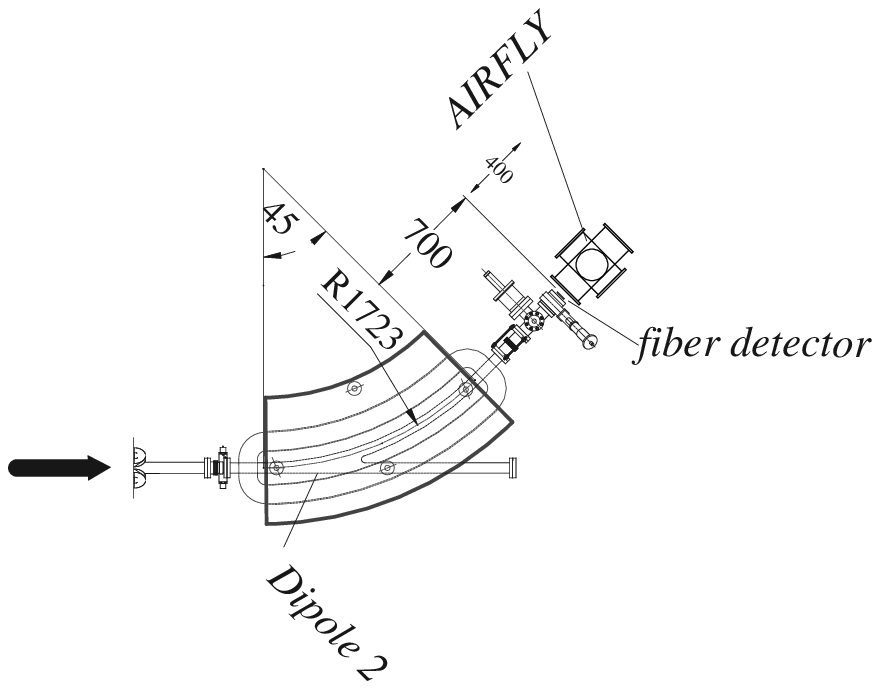,width=0.5\textwidth}}
        \caption{The scintillating fiber detector is usually placed at the
        the $45^\circ$ beam exit (with the final dipole on), dedicated to user setups.}
        \label{fig:btfplan}
\end{figure}

In all the following BTF running periods, both the $x$ and $y$ views of the
detector were mounted, on their mechanical support, again, right at the exit of the
BTF beam for the users (Fig.~\ref{fig:btfplan}).

\subsection{Detector readout}

The 16 analog signals of the MAPMT pixels are splitted in two by a passive splitter board.
The suitably delayed signals are digitized by a CAEN V792 charge integrating ADC, 
0.25 pC/count. The gate signal, 100 ns long, is generated starting from a reference signal
generated by the LINAC timing circuit for each
beam pulse. The analog signals are also digitized by means of a low
threshold discriminator (CAEN V814), with a typical threshold of  35 mV$/50\Omega$,
and the time of each channel is measured by means of a CAEN V775 VME TDC, 35 ps/count.
The VME controller, a VMIC 7740 Pentium III CPU with Tundra VME-PCI chip, runs Red Hat Linux 7.2 and a 
LabVIEW 6.1 DAQ program. Further details on the DAQ can be found in Ref.~\cite{bib:btfnote}.

\section{Experimental results}
\label{sec:results}

The beam profile is measured starting from the pulse height measured in each of
the 2$\times$16 MAPMT pixels, by means of the charge integrating ADC (once
having subtracted the pedestal).
This pulse height infact should be proportional to the charge deposited in each
fibers bundle. 

Since the detector is very thin, $\approx$ 2\% of a radiation length,
each electron transversing it should have a well defined average energy loss,
of course with fluctuations following the typical Landau distribution, so
that the average charge in a pixel should be proportional to the total number of
electrons crossing the corresponding bundle.

In order to check this, the pulse height in the fiber detector has been
measured in a run at low electron multiplicity, i.e. in single
electron mode. Since the number of particles in
the beam follows the Poisson statistics, the pulse height in the calorimeter,
placed downstream of the fiber detector, has been used to separate events
with none, one or two electrons crossing the detector.
\begin{figure}[tbhp]
  \centerline{\epsfig{file=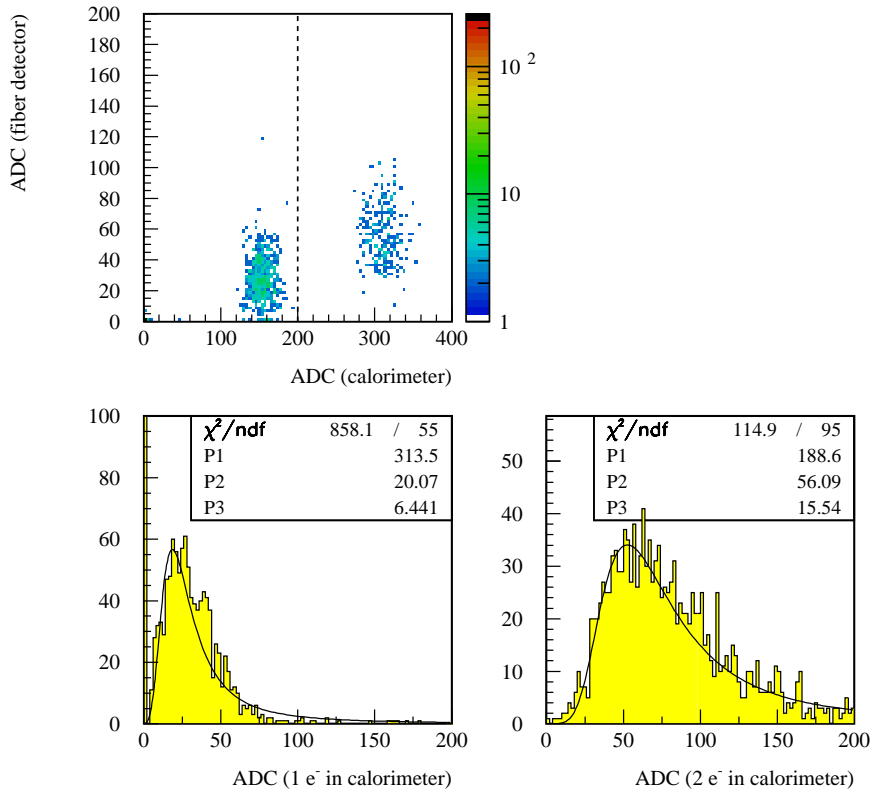,width=0.8\textwidth} }
        \caption{Top: correlation of the pulse height in the fiber detector (sum of 
          the 16 MAPMT channels pedestal subtracted) with the pulse height
          in the calorimeter (see text). 
          Bottom: the fiber detector pulse height spectrum for events 
          with one (left plot) or two electrons (right plot), fits
          to the Landau function are also shown.}
        \label{fig:corrcalo}
\end{figure}

In Fig.~\ref{fig:corrcalo} the pulse height in one view of the detector, 
summed over all the fibers, is shown as a function of the pulse
height in the calorimeter, for a run at 493 MeV, with an average
multiplicity of 1 particle/pulse. Using the total energy deposited in
the calorimeter, placed $\approx$ a 1.5 m distance from the beam exit window, 
at a distance of 1.1 m from the fiber detector,
one can separate one and two electrons events, the pulse height spectrum
in the fibers clearly follows a Landau distribution.

The shape and the position of the beam spot can then be determined by
the charge-weighted distribution of events in the 16 ``fingers'', 3 mm wide, 
of fibers constituting each of the two views of the detector.
Since the BTF is usually operated at the maximum repetition rate of 49 bunches/s,
an accurate measurement of the beam spot can be achieved already accumulating
a few seconds of beam, even at the lowest particle multiplicity, i.e. in 
single electron mode.

A possible problem in building the charge-weighted distribution can be the
non-uniformity of the response of each ``finger'', when crossed by a single
electron. This can be due to a number of reasons; the main one
in our case is the non-optimal coupling of the fiber bundles to the 
surface of the MAPMT cathode, since no optical glue has been used, and to 
the possibility of breaking one or more fibers when inserting the 
bundles in the grooved PVC mask.
\begin{figure}[tbhp]
  \centerline{\epsfig{file=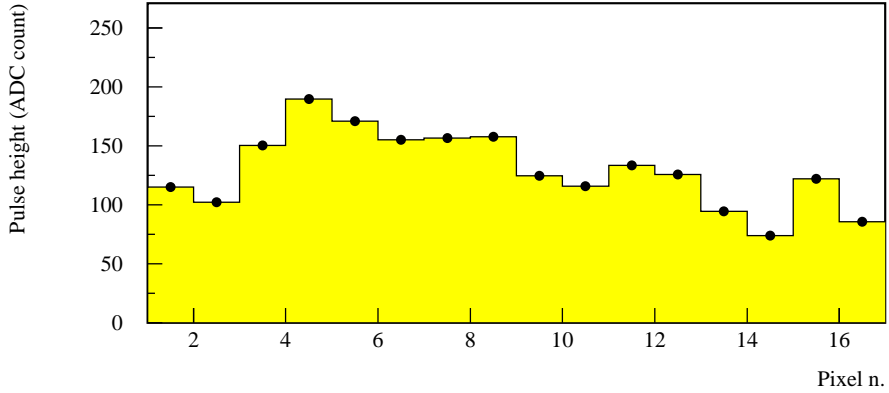,width=0.8\textwidth} }
        \caption{Pulse height of the 16 MAPMT channels (pedestal subtracted);
          in order to have an approximately uniform illumination (the
          detector was arranged with the fibers running in the horizontal
          direction, with the beam defocussed in the vertical plane).
          This spectrum has been used for the pixel equalization; the yield differences
          between the pixels are mainly due to broken/inefficient fibers (as shown by 
          the groups of pixels with approximately the same yield).}
        \label{fig:norma}
\end{figure}

In order to estimate the uniformity of response of the pixels in each
view of the detector, dedicated runs with intermediate multiplicity
and a very defocussed beam have been analyzed. In this data, the beam
is constituted by a few tens of electrons distributed over an area,
at the beam exit window, of $\approx$ 55$\times$25 mm$^2$, so that
the surface of the fiber detector is almost uniformely illuminated.
In this condition we could expect an uniform response from all the
16 channels in one view. In Fig.~\ref{fig:norma} the pulse height
distribution (pedestal subtracted) of all the channels of one view is shown.
There are evident yield differences between the pixels; since there
are anyhow groups of pixel with approximately the same response, the
differences can mainly be interpreted as due to broken or inefficient 
fibers in the single bundles.

The normalized output in the defocussed beam conditions are used to correct 
the relative yield of each pixel, for both views of the detector.

\begin{figure}[tbhp]
  \centerline{\epsfig{file=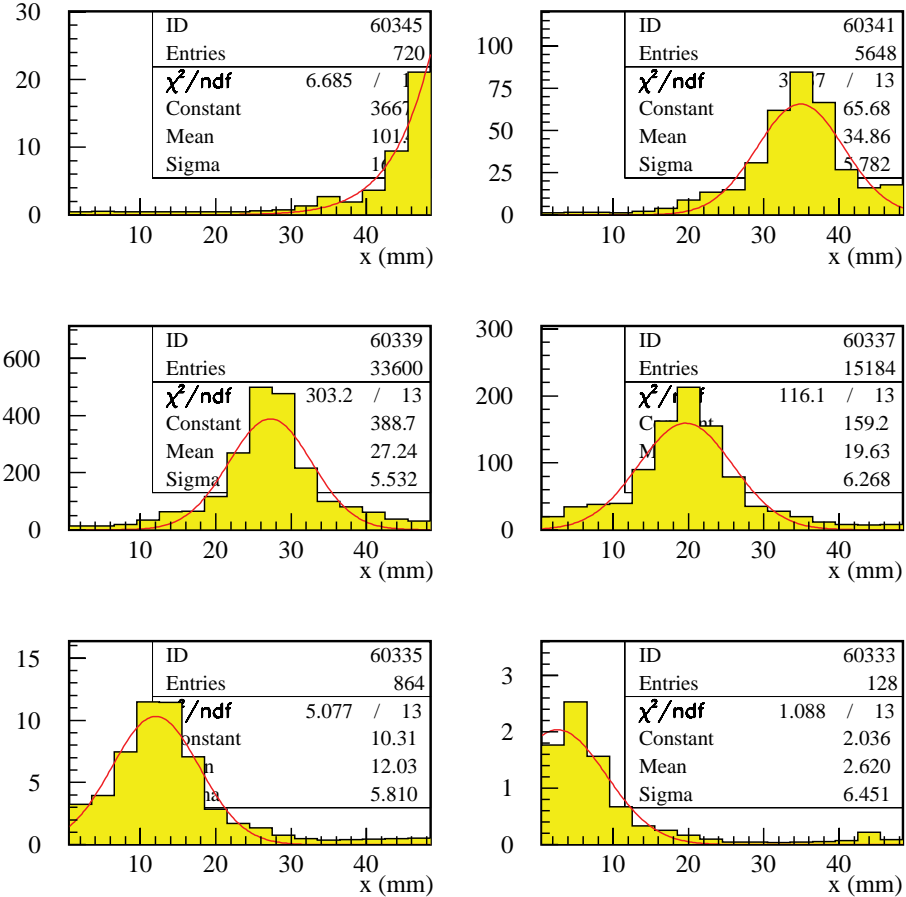,width=0.8\textwidth} }
        \caption{Charge-weighted profile for six different settings of
          the final bending, corresponding to a scan in the horizontal 
          direction (along the $x$ axis).}
        \label{fig:xscan}
\end{figure}

In order to check the functionality of the first of the two detector planes,
some runs were taken moving the beam along the horizontal axis (with the
detector arranged with the fibers running along the vertical axis), using a 
standard focussed beam setting of the BTF, yielding 493 MeV electrons, on average
1 electron per pulse, with a spot size of few mm. The horizontal position of
the beam has been changed by changing the current of the last bending magnet
in small steps (2 A on a nominal setting of 335A), thus resulting in a 
nominal deviation of $\approx$7 mm at the beam exit window. 
The results of this scan
in the horizontal direction are shown in Fig.~\ref{fig:xscan}.

The shape of the charge-weighted profile is well reproduced by a
Gaussian. By fitting the distribution, the mean values well
reproduce the expected beam deviation. Concerning the width
of the distribution, different contribution can be identified:
\begin{itemize}
\item the intrinsic beam spot size; this has been measured with
much more accurate detectors (AGILE Silicon Tracker) to be fairly Gaussian 
with $\sigma\simeq 2.2$ mm in both dimensions (with the
optimized focussed beam setting we were using, of 493 MeV single 
electron beam); this includes the natural width of the beam plus
the multiple scattering contribution on the thin aluminum exit window; 
\item the resolution of the fiber detector;
\item if the detector is placed at some distance from the exit window,
the contribution of the multiple scattering in air.
\end{itemize}

Since the multiple scattering is momentum-dependant, this contribution
can be disentangled by performing the beam spot measurement at different
beam energies; of course this is possible only if the beam spot size is
not significantly changed by changing the BTF line settings for the
different momentum settings.

\begin{figure}[tbhp]
  \centerline{\epsfig{file=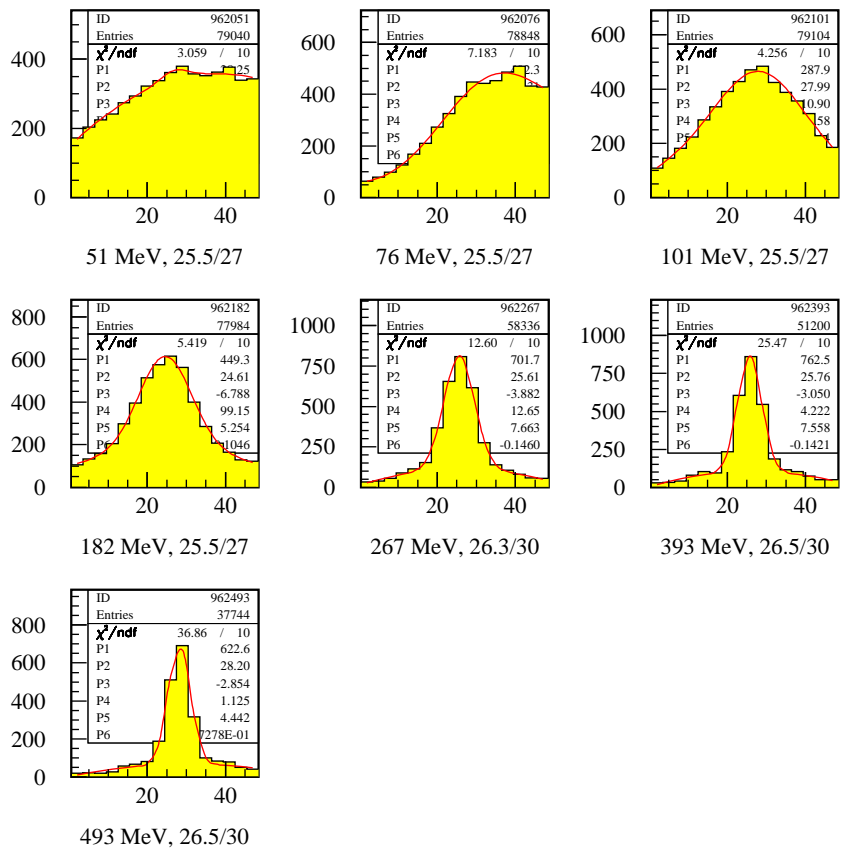,width=0.8\textwidth} }
     \caption{Charge-weighted profile for different BTF beam energies:
          the (horizontal) beam spot crearly increases at low energy
          due to the contribution of the multiple scattering
          (mainly due to the 1.5 mm thick aluminum window).}
        \label{fig:escana}
\end{figure}

The measured beam profiles, only for the horizontal view of the
fiber detector, in a wide range of electron energy, with an optimal
focussing of the beam and with horizontal collimators almost
completely closed (thus with an approximately unchanged intrinsic
horizontal beam size), are shown in Fig.~\ref{fig:escana}.

\begin{figure}[tbhp]
  \centerline{\epsfig{file=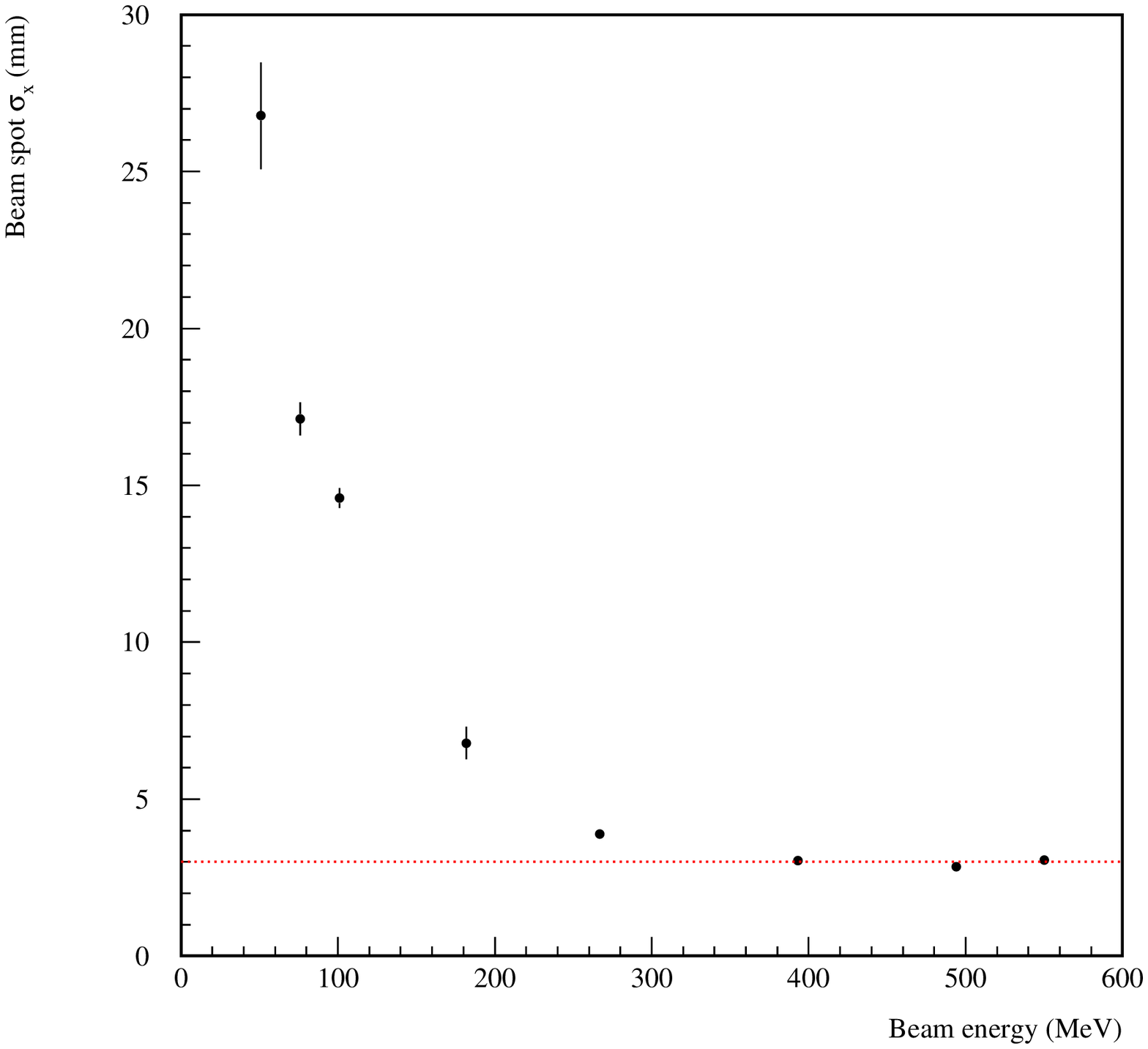,width=0.8\textwidth} }
     \caption{Horizontal beam spot size as a function of the beam energy:
          the behaviour is consistent with the $1/(\beta c p)$
          dependance of the multiple scattering RMS angle at low energy, 
          while approaches a constant value at higher energies, given by
          the contributions of the beam divergence and the fiber detector spatial 
          resolution.}
        \label{fig:escanb}
\end{figure}

The $\sigma$ values for the Gaussian fits of the profiles of Fig.~\ref{fig:escana}
are shown in Fig.~\ref{fig:escanb} as a function of the beam energy. 
The expected $1/(\beta c p)$ dependance for the multiple scattering contribution
can be observed at low energies, while the measured beam spot approches a
constant value at energies above 400 MeV, of $\approx$ 3 mm. Two
main effects contribute to the plateau value: the spatial resolution of
the fiber detector and the widening of the beam spot due to the intrinsic 
divergence, that is not negligible even if the beam is strongly collimated in 
the BTF line.

%%%%%%%%%%%%%%%%%%%%%%%%%%%%%%%%%%%%%%%%%%%%%%%%%%%%%%%%%%%%%%%%%%%%%%%%%%%%%%%%%%%%%%%%%%%%%%

\begin{figure}[tbhp]
  \centerline{\epsfig{file=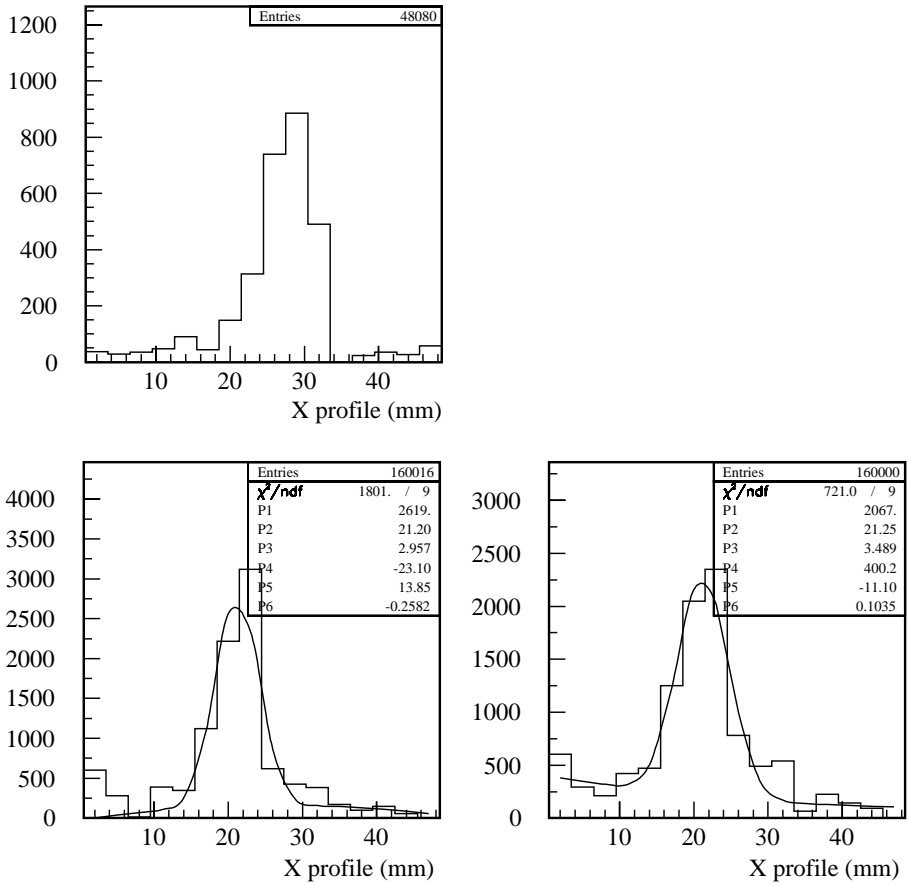,width=0.8\textwidth} }
     \caption{Horizontal beam spot size measured for different beam
        intensities and very similar beam focussing and collimation
        conditions: top, $\approx$ 2 particles/pulse; bottom left, $\approx$150 particles/pulse; 
        bottom right, $\approx$600 particles/pulse.}
        \label{fig:sizevsn}
\end{figure}

Another important point is the possibility of using the fiber detector
also at intermediate beam intensities, i.e. between tens and thousands
of particles per pulse, without any significant performance loss.
In order to check this, the beam spot has been monitored for a focussed beam
while increasing the beam intensity. No significant difference in the measured
beam spot can be found up to several hundreds particles per bunch, as shown
in the example in Fig. \ref{fig:sizevsn}, where the horizontal profile is
shown for a single particle beam, for $\approx$ 150, up to more than
600 particles/pulse.

\section{Conclusions}
The scintillating fiber detector has been built in order to have 
a simple, robust, easy to build and to manage beam spot monitor, capable of
a millimetric resolution in a wide range of the BTF beam operating
parameters (energy, intensity, focussing). 
The granularity of the readout has been sacrified to the possibility
of having as little as 32 ADC channels, so that they are all housed
in a single 6U VME board, with only two 16-channels multianode
photomultipliers, and to the possibility of having a good light
yield even with only one electron crossing the fibers, with a relatively
thin detector (only four layers per view).

However, the detector performed very well during the 2003 data taking, and
allowed to continuously monitor the position and
the size of the beam, with an accuracy of $\approx$ 2 mm, in both views. 

\begin{figure}[tbhp]
  \centerline{\epsfig{file=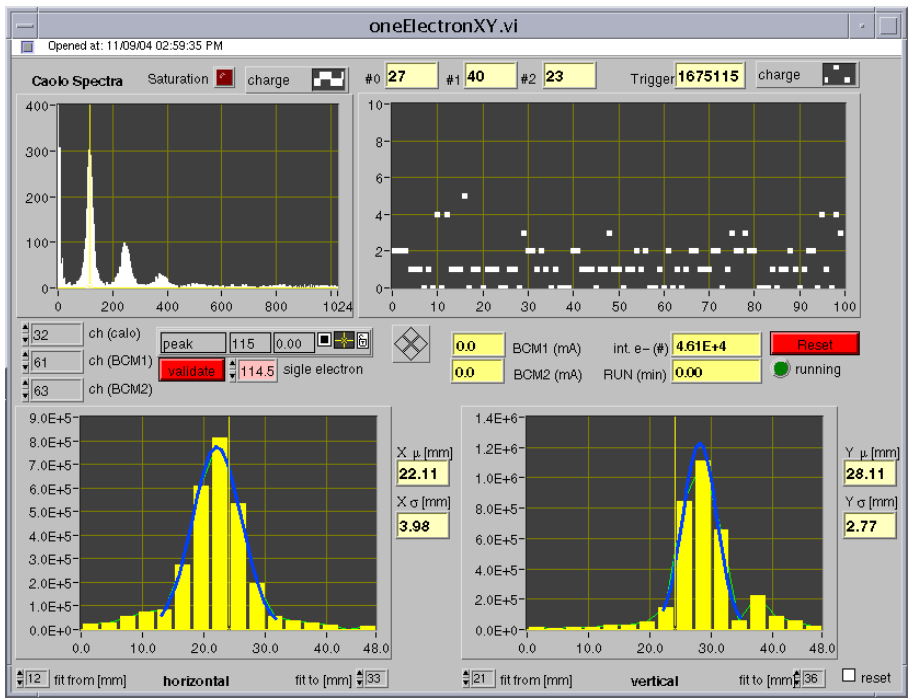,width=0.8\textwidth} }
     \caption{The horizontal and vertical beam profiles measured by means
       of the fiber detector have been integrated in the \DAF\ BTF control 
       system,
       allowing an online monitoring of the position and size of the beam spot.}
        \label{fig:tool}
\end{figure}

Since
a few seconds of data taking, at the maximum repetition rate of 49 pulses/s,
are sufficient to get a satifactory beam profile, the horizontal and vertical
charge-weighted histogram have been integrated in the \DAF\ BTF contol
system. The shape of the beam in both $x$ and $y$ views can then be viewed online
in the main panel of the BTF control system, as shown in Fig.~\ref{fig:tool}.
This demonstrated to be very useful tool, both in the beam optimization phases, 
and for driving the beam onto the users experimental apparata with a good accuracy.

Most importantly, the detector is effective in a wide range of beam intensities,
being efficient in single particle mode, and in the $10^3$ particles/pulse range
without any significant loss of resolution.

%%%%%%%%%%%%%%%%%%%%%%%%%%%%%%%%%%%%%%%%%%%%%%%%%%%%%%%%%%%%%%%%%%%%%%%%%%%%%%%%%%%%%%%%%%%%%%
%%%%%%%%%%%%%%%%%%%%%%%%%%%%%%%%%%%%%%%%%%%%%%%%%%%%%%%%%%%%%%%%%%%%%%%%%%%%%%%%%%%%%%%%%%%%%%
\clearpage
\section*{Acknowledgements}

We thank U.~Frasacco for the MAPMT cabling, G. Ferretti for the
MAPMT mask construction, R.~Clementi and
R.~Zarlenga for the scintillating fibers polishing, M. Sperati for the mechanical support.

We are grateful to S.~Miscetti for the useful suggestions and for providing us the MAPMT.

We deeply thank P.~Privitera and all the AIRFLY group for the precious collaboration 
during the data taking at the BTF.

Work partially supported by TARI contract HPRI-CT-1999-00088.

%%%%%%%%%%%%%%%%%%%%%%%%%%%%%%%%%%%%%%%%%%%%%%%%%%%%%%%%%%%%%%%%%%%%%%%%%%%%%%%%%%%%%%%%%%%%%%
%%%%%%%%%%%%%%%%%%%%%%%%%%%%%%%%%%%%%%%%%%%%%%%%%%%%%%%%%%%%%%%%%%%%%%%%%%%%%%%%%%%%%%%%%%%%%%

\end{document}